\documentclass[nofootinbib,twocolumn,showpacs,preprintnumbers,pre,aps]{revtex4-1}

\usepackage[dvipdfmx]{graphicx}
\usepackage{bm}
\usepackage{amsmath}
\usepackage{amssymb}
\usepackage{color}
\usepackage{here}
\usepackage{siunitx}

\begin{document}

\title{Irreversibility and entropy production of a thermally driven micromachine}

\author{Isamu Sou}

\author{Yuto Hosaka}

\author{Kento Yasuda}

\author{Shigeyuki Komura}
\email{komura@tmu.ac.jp}

\affiliation{
Department of Chemistry, Graduate School of Science,
Tokyo Metropolitan University, Tokyo 192-0397, Japan}


\begin{abstract}
We discuss the non-equilibrium properties of a thermally driven micromachine consisting 
of three spheres which are in equilibrium with independent heat baths characterized by 
different temperatures.
Within the framework of a linear stochastic Langevin description, we calculate the 
time-dependent average irreversibility that takes a maximum value for a finite time.
This time scale is roughly set by the spring relaxation time.
The steady-state average entropy production rate is obtained in terms of the 
temperatures and the friction coefficients of the spheres. 
The average entropy production rate depends on thermal and/or mechanical asymmetry
of a three-sphere micromachine.
We also obtain the center of mass diffusion coefficient of a thermally driven three-sphere 
micromachine as a function of different temperatures and friction coefficients.
With the results of the total entropy production rate and the diffusion coefficient, we finally   
discuss the efficiency of a thermally driven micromachine.
\end{abstract}

\maketitle

\section{Introduction}
\label{sec:introduction}

Microswimmers are tiny machines that swim in a fluid and they are expected to be used in microfluidics 
and microsystems~\cite{Lauga09}.
By transforming chemical energy into mechanical energy, microswimmers change their shape 
and move efficiently in viscous environments.
According to Purcell's scallop theorem, reciprocal body motion cannot be 
used for locomotion in a Newtonian fluid~\cite{Purcell77,Lauga11}.
As one of the simplest models showing non-reciprocal body motion, 
Najafi and Golestanian proposed a three-sphere swimmer~\cite{Golestanian04,Golestanian08}, 
in which three in-line spheres are linked by two arms of varying length.
Such a swimmer has been experimentally realized by using colloidal beads manipulated by 
optical tweezers~\cite{Leoni09} or by controlling ferromagnetic particles at an air-water 
interface~\cite{Grosjean16,Grosjean18}.

\begin{figure}[bth]
\begin{center}
\includegraphics[scale=0.35]{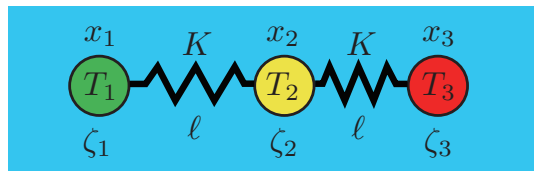}
\end{center}
\caption{(Color online)　Thermally driven elastic three-sphere micromachine.
Three spheres are connected by two identical harmonic springs characterized by the elastic 
constant $K$ and the natural length $\ell$.
The time-dependent positions of the spheres are denoted by $x_i(t)$ ($i=1, 2, 3$) in a 
one-dimensional coordinate system, and $\zeta_i$ is the friction coefficient for $i$-th sphere.
The three spheres are in equilibrium with independent heat baths having different temperatures $T_i$.
}
\label{model}
\end{figure}

Recently, the present authors have proposed a generalized three-sphere microswimmer model 
in which the spheres are connected by two harmonic springs, i.e., an elastic 
microswimmer~\cite{Yasuda17,Kuroda19}.  
Later, our model was further extended to a thermally driven elastic microswimmer~\cite{Hosaka17}, 
suggesting a new mechanism for locomotion that is purely induced by thermal fluctuations
without any external forcing.
The key setting of the model is that the three spheres are in equilibrium with independent 
heat baths characterized by different temperatures (as described later in Fig.~\ref{model}).
For this thermally driven three-sphere micromachine, the average velocity was calculated to 
be~\cite{Hosaka17} 
\begin{align}
\langle V\rangle=\frac{k_{\rm B}(T_3-T_1)}{96\pi\eta\ell^2},
\label{Vthermal}
\end{align}
where $k_{\rm B}$ is the Boltzmann constant, $T_1$ and $T_3$ are the temperatures 
of the first and the third spheres (see Fig.~\ref{model}), $\eta$ is the viscosity of the 
surrounding fluid, and $\ell$ is the natural length of the two springs.
We have shown that a combination of heat transfer and hydrodynamic interactions 
among the spheres leads to directional locomotion in a steady-state, which can be described 
in terms of ``stochastic energetics''~\cite{Sekimoto97,Sekimoto98,SekimotoBook}.

Systems in thermodynamic equilibrium obey detailed balance meaning that transition rates between 
any two microscopic states are pairwise balanced~\cite{KampenBook,RiskenBook}.
For non-equilibrium steady-state situations, however, detailed balance is broken and 
a probability flux loop exists in a configuration phase 
space~\cite{Battle16,Gladrow16,Gladrow17,Gnesotto18}.
Recently, the present authors discussed the non-equilibrium steady-state probability distribution function 
of a thermally driven three-sphere micromachine and calculated its probability flux in the corresponding 
configuration space~\cite{Sou19}. 
The resulting probability flux can be expressed in terms of a frequency matrix to characterize 
a non-equilibrium steady-state~\cite{Weiss03,Weiss07}.
Importantly, we have obtained a linear relation between the eigenvalue of the frequency matrix and the 
average velocity of a thermally driven micromachine~\cite{Sou19}.

Since our model micromachine offers a new type of thermal ratchet~\cite{Hosaka17}, a more 
detailed analysis based on non-equilibrium statistical mechanics is required in order to elucidate the 
physical mechanism for the locomotion of a thermally driven three-sphere micromachine.
One of the important quantities to measure the statistical irreversibility of a non-equilibrium process 
is the entropy production rate~\cite{Seifert05,Jarzynski11,Seifert12}.
In this paper, within the framework of a linear stochastic Langevin description, we calculate the 
time-dependent average irreversibility of a thermally driven three-sphere micromachine in the 
absence of hydrodynamic interactions.
The average irreversibility is important because its initial growth rate gives the average entropy 
production rate.
Using this fact, we explicitly obtain the average entropy production rate for a thermally driven 
three-sphere micromachine, and express it in terms of different temperatures and friction coefficients.
We examine in detail how the entropy production rate depends on the asymmetry of the 
temperatures and/or the friction coefficients of a three-sphere micromachine.

Although a micromachine does not exhibit any directional motion in the absence of hydrodynamic 
interactions~\cite{Sou19}, it undergoes a thermal Brownian motion.
In addition to the above mentioned non-equilibrium quantities, we also calculate the center of 
\textit{mass} diffusion coefficient of a thermally driven three-sphere micromachine
(even though we neglect the inertia of the spheres).
Using the results of the diffusion coefficient for a dimer and trimer, we predict a simple and useful 
expression for the center of mass diffusion coefficient of an elastic $n$-sphere micromachine.
We consider that our expression is useful because the center of \textit{mass} diffusion is easier 
to measure than the center of \textit{friction} diffusion~\cite{Grosberg15}.
Finally, we obtain the total entropy production rate including the center of mass motion, and 
further discuss the efficiency of a thermally driven three-sphere micromachine.
Our result shows that the efficiency becomes larger for specific combinations of the temperatures.

In the next section, we briefly review the framework of a linear Langevin model and describe 
how the average irreversibility and the entropy production rate are obtained in general. 
In Sec.~\ref{sec:elastic}, we explain our model of a thermally driven three-sphere micromachine 
by introducing the coupled Langevin equations for the two spring lengths~\cite{Sou19}.
In Secs.~\ref{sec:irreversibility} and \ref{sec:entropy}, we explicitly calculate the average 
irreversibility and the average entropy production rate, respectively, for a thermally driven 
three-sphere micromachine.
The center of mass diffusion coefficient of a thermally driven micromachine is given 
in Sec.~\ref{sec:diffusion}.
In Sec.~\ref{sec:efficiency}, after calculating the total entropy production rate, 
we discuss the efficiency of a micromachine.
Finally, a summary of our work and some discussion are given in Sec.~\ref{sec:discussion}.

\section{Linear Langevin system}
\label{sec:linlangevin}

In this section, we briefly review Ref.~\cite{Weiss07} and pick up the important results for
our calculation.
Let us consider a linear stochastic Langevin equation given by~\cite{ZwanzigBook}
\begin{align}
\frac{d{\bm r}(t)}{d t} = {\bf A} {\bm r}(t) + {\bf F} {\bm \xi}(t), 
\label{langevingeneral}
\end{align}
where ${\bm r}$ is the $N$-dimensional state vector of real numbers,
${\bf A}$ is an $N \times N$ real matrix representing the linear deterministic dynamics, 
${\bf F}$ is an $N \times N$ real matrix representing the noise forcing.
We require that all eigenvalues of ${\bf A}$ have a negative real part so that the system 
will eventually reach a steady-state. 
In Eq.~(\ref{langevingeneral}), ${\bm \xi}$ is $N$-dimensional Gaussian white noise satisfying 
the statistical properties
\begin{align}
& \langle {\bm \xi}(t) \rangle =0,
\\
& \langle {\bm \xi}(t) {\bm \xi}(t')^{\mathsf T} \rangle = {\bf B} \delta(t - t'), 
\end{align}
where ${\bf B}$ is an $N \times N$ matrix representing the variance of noise,  
and superscript ${\mathsf T}$ represents the transpose.
Then the diffusion matrix ${\mathbf D}$ is obtained by 
\begin{align}
{\mathbf D}  = \frac{1}{2} {\mathbf F} {\mathbf B} {\mathbf F}^{\mathsf T}.
\label{Dmatrix}
\end{align}

Using the above Langevin model, we discuss the probability of observing trajectory segments. 
For any two states ${\bm r}_{0}$ and ${\bm r}_{1}$, we consider the trajectory probability
$p({\bm r}_{0},{\bm r}_{1},t)$ as the probability of finding a trajectory segment within 
the long trajectory which begins at ${\bm r}_{0}$ and ends at ${\bm r}_{1}$ a time $t$ later.
Such a trajectory probability can be expressed as 
\begin{align}
p ({\bm r}_{0}, {\bm r}_{1},t) = p({\bm r}_{1},t | {\bm r}_{0})p_{0}({\bm r}_{0}), 
\label{pr0r1}
\end{align}
where $p({\bm r}_{1},t | {\bm r}_{0})$ is the transition probability of finding the system in state 
${\bm r}_{1}$ conditioned on the system being in state ${\bm r}_{0}$ a time $t$ earlier, and 
$p_{0}({\bm r}_{0})$ is the steady-state probability of finding the system in state ${\bm r}_{0}$.
Similarly, the time-reversed trajectory segment, one starting at ${\bm r}_{1}$ and ending at ${\bm r}_{0}$,
has a probability 
\begin{align}
p ({\bm r}_{1}, {\bm r}_{0},t) = p({\bm r}_{0},t | {\bm r}_{1})p_{0}({\bm r}_{1}).
\label{pr1r0}
\end{align}

The irreversibility $\sigma({\bm r}_{0},{\bm r}_{1},t)$ of a trajectory segment with initial state 
${\bm r}_{0}$ and final state ${\bm r}_{1}$ is defined by~\cite{Weiss07}
\begin{align}
\sigma({\bm r}_{0},{\bm r}_{1},t) = \ln{\frac{p({\bm r}_{0},{\bm r}_{1},t)}{p({\bm r}_{1},{\bm r}_{0},t)}}.
\label{irreversibility}
\end{align}
The system is reversible when $\sigma=0$, while forward and reverse trajectories are distinguishable
when $\sigma \neq 0$.
By introducing the probability $P(\sigma)$ of finding a trajectory segment with irreversibility $\sigma$, 
the fluctuation theorem can be expressed as~\cite{Weiss07,Seifert05,Jarzynski11,Seifert12}
\begin{align}
\frac{P(\sigma)}{P(-\sigma)}=e^{\sigma}.
\label{fluctuation}
\end{align}
Although this fluctuation theorem gives a constraint on $P(\sigma)$, it does not 
completely fix its functional form.
How to obtain $P(\sigma)$ for a linear Langevin system is separately explained in Appendix~\ref{app:irrpdf}.

Because the model in Eq.~(\ref{langevingeneral}) is linear with additive Gaussian white noise, 
the probabilities in Eqs.~(\ref{pr0r1}) and (\ref{pr1r0}) are also Gaussian, and they can be 
written in terms of the covariance of the dynamics.
Weiss showed that the time-dependent average irreversibility $\langle \sigma (t) \rangle$ is 
given by~\cite{Weiss07} 
\begin{align}
\langle \sigma (t) \rangle = {\rm Tr} \left[{\bf C}_0 {\bf C}_t^{-1} 
({\bf I} - e^{2 {\bf A} t}) - {\bf I} \right], 
\label{meanirr}
\end{align}
where ${\bf I}$ is the $N \times N$ identity matrix.
In the above, ${\bf C}_0$ is the steady-state covariance matrix satisfying the Lyapunov 
equation~\cite{Netz18}
\begin{align}
{\bf A} {\bf C}_0 + {\bf C}_0 {\bf A}^{\mathsf T} + 2 {\bf D} = 0,
\label{lyapunov}
\end{align}
which can be regarded as the fluctuation-dissipation relation~\cite{Weiss03}.
Moreover, ${\bf C}_t$ is the time-dependent covariance matrix of the transition probability 
and is given by  
\begin{align}
{\bf C}_{t} & = 2 \int_{0}^{t} ds \, e^{{\bf A}(t-s)} {\bf D} e^{{\bf A}^{\mathsf T} (t-s)}
\nonumber \\
& = {\bf C}_{0} - e^{{\bf A}t} {\bf C}_{0} e^{{\bf A}^{\mathsf T} t}.
\label{Cta}
\end{align}
Notice that ${\bf C}_0$ and ${\bf C}_{t}$ are related by 
${\bf C}_0 = \lim_{t \rightarrow \infty} {\bf C}_{t}$.

In Ref.~\cite{Weiss07}, it was further shown that the steady-state average entropy production 
rate $\langle \dot{\sigma} \rangle$ is given by the zero-time growth rate of the average 
irreversibility in Eq.~(\ref{meanirr}), i.e.,
\begin{align}
\langle \dot{\sigma} \rangle & = \left. \frac{d \langle \sigma (t) \rangle}{dt} \right |_{t=0}
= {\rm Tr} \left[{\bf A} {\bf G} \right].
\label{entproorig}
\end{align}
Here, ${\bf G}$ is the dimensionless gain matrix defined by 
\begin{align}
{\bf G}=-({\bf A} {\bf C}_{0} {\bf D}^{-1} + {\bf I}).
\label{gain}
\end{align}
The gain matrix ${\bf G}$ is directly related to the violation of detailed balance~\cite{Weiss03}, and 
the product ${\bf A} {\bf G}$ in Eq.~(\ref{entproorig}) measures the noise amplification per unit 
time~\cite{Weiss07}.

It is worth mentioning that the same average entropy production rate in Eq.~(\ref{entproorig})
can also be obtained directly from the steady-state probability flux~\cite{Weiss07,Li19}.
In fact, the latter approach is more standard~\cite{Chernyak06}.
For a three-sphere micromachine, the steady-state probability flux was calculated 
in our previous paper~\cite{Sou19}.

\section{Thermally driven three-sphere micromachine}
\label{sec:elastic}

In this section, we explain the model of a thermally driven elastic micromachine that has been introduced 
in our previous studies~\cite{Hosaka17,Sou19}.
As schematically shown in Fig.~\ref{model}, the model consists of three hard spheres connected 
by two harmonic springs.
We assume that the two springs are identical, and the common spring constant and the natural length 
are given by $K$ and $\ell$, respectively.
The positions of the three spheres in a one-dimensional coordinate system is defined as 
$x_i(t)$ ($i=1, 2, 3$).

Most importantly, we consider a situation where the three spheres are in thermal equilibrium 
with independent heat baths at temperatures $T_i$~\cite{Hosaka17,Sou19}.
When these temperatures are different, the system is driven out of equilibrium 
because heat flux from a hotter sphere to a colder one is generated. 
The Langevin equations of motion of the three spheres are given by 
\begin{align}
\frac{d x_1}{dt} &= \frac{K}{\zeta_1}(x_2-x_1-\ell)+\left(\frac{2T_1}{\zeta_1}\right)^{1/2}\xi_1,
\label{x1dot} \\
\frac{d x_2}{dt} &= - \frac{K}{\zeta_2}(x_2-x_1-\ell)+\frac{K}{\zeta_2}(x_3-x_2-\ell)
\nonumber \\
& +\left( \frac{2T_2}{\zeta_2} \right)^{1/2} \xi_2, 
\label{x2dot} \\
\frac{d x_3}{dt} &=- \frac{K}{\zeta_3}(x_3-x_2-\ell)+\left( \frac{2T_3}{\zeta_3} \right)^{1/2} \xi_3,
\label{x3dot}
\end{align}
where $\zeta_i$ is the friction coefficient for $i$-th sphere, and the Boltzmann constant $k_{\rm B}$
is set to unity hereafter (except later in Sec.~\ref{sec:efficiency}).
As we discuss later, the friction coefficient is generally proportional to the size of the sphere.
Furthermore, $\xi_i(t)$ is a zero mean and unit variance Gaussian white noise, 
independent for all the spheres: 
\begin{align}
& \langle \xi_i(t)\rangle  =0,
\\
& \langle \xi_i(t)\xi_j(t') \rangle  = \delta_{ij}\delta{(t-t')}.
\end{align}

In contrast to Ref.~\cite{Hosaka17}, we do not consider hydrodynamic interactions acting 
between the different spheres.
Although the locomotion of a micromachine is not discussed in this paper (except 
in Sec.~\ref{sec:efficiency}), we use a term ``micromachine" for the above three-sphere system
because it can undergo a directional motion in the presence of hydrodynamics interactions~\cite{Hosaka17}.
The effects of hydrodynamic interactions will be separately discussed in Sec.~\ref{sec:discussion}.

To describe the configuration of a micromachine, it is convenient to introduce the following two 
spring extensions with respect to $\ell$:
\begin{align}
r_{12}=x_2-x_1-\ell,~~~~~r_{23}=x_3-x_2-\ell.
\label{r12r23}
\end{align}
From Eqs.~(\ref{x1dot})--(\ref{x3dot}), we obtain the reduced Langevin equations for $r_{12}(t)$ and 
$r_{23}(t)$ as~\cite{Grosberg15}
\begin{align}
\frac{d r_{12}}{dt} &= -\frac{K}{\zeta_{12}} r_{12} + \frac{K}{\zeta_2} r_{23} + 
\left( \frac{2T_{12}}{\zeta_{12}} \right)^{1/2} \xi_{12}, 
\label{eqr12} \\
\frac{d r_{23}}{dt} &=  \frac{K}{\zeta_2}r_{12} - \frac{K}{\zeta_{23}} r_{23} +
\left( \frac{2T_{23}}{\zeta_{23}} \right)^{1/2} \xi_{23}.
\label{eqr23}
\end{align}
Here we have introduced the relevant effective friction coefficient 
\begin{align}
\zeta_{ij} = \frac{\zeta_i\zeta_j}{\zeta_i + \zeta_j},
\label{zetaij}
\end{align}
and the friction-weighted average temperature
\begin{align}
T_{ij} = \frac{\zeta_j T_i + \zeta_i T_j}{\zeta_i + \zeta_j}. 
\label{Tij}
\end{align}
The definition of the effective temperature $T_{ij}$ arises from the requirement 
that the newly introduced noises $\xi_{12}(t)$ and $\xi_{23}(t)$ in 
Eqs.~(\ref{eqr12}) and (\ref{eqr23}), respectively, satisfy the following statistical properties:
\begin{align}
& \langle \xi_{12}(t)\rangle= \langle \xi_{23}(t)\rangle = 0,
\\
& \langle \xi_{12}(t)\xi_{12}(t')\rangle=  \delta(t-t'), 
\label{xi12xi12}
\\
& \langle \xi_{23}(t)\xi_{23}(t')\rangle=  \delta(t-t'), 
\label{xi23xi23}
\\
& \langle \xi_{12}(t)\xi_{23}(t')\rangle= 
- \frac{T_{2}}{\zeta_{2}}\left( \frac{\zeta_{12}\zeta_{23}}{T_{12}T_{23}}\right)^{1/2} \delta(t - t').
\label{xi12xi23}
\end{align}

The reduced Langevin equations in Eqs.~(\ref{eqr12}) and (\ref{eqr23}) can be conveniently represented 
in the matrix form of Eq.~(\ref{langevingeneral}). 
Using the notations for the two-dimensional vectors
${\bm r} = (r_{12}, r_{23})^{\mathsf T}$ and ${\bm \xi} = (\xi_{12} , \xi_{23})^{\mathsf T}$, 
the $2\times 2$ matrices ${\bf A}$, ${\bf F}$, and ${\bf B}$ are now given by  
\begin{align}
{\mathbf A} = 
\begin{pmatrix} 
-K/\zeta_{12} &  K/\zeta_2 
\\
K/\zeta_2 & - K/\zeta_{23}
\end{pmatrix},
\label{Amatrixthree}
\end{align}
\begin{align}
{\mathbf F} = 
\begin{pmatrix} 
\left( \dfrac{2T_{12}}{\zeta_{12}} \right)^{1/2} &  0
\\
0 & \left( \dfrac{2T_{23}}{\zeta_{23}} \right)^{1/2}
\end{pmatrix},
\label{Fmatrixthree}
\end{align}
and 
\begin{align}
{\bf B} = \begin{pmatrix} 
1 &  - \dfrac{T_{2}}{\zeta_{2}}\left( \dfrac{\zeta_{12}\zeta_{23}}{T_{12}T_{23}}\right)^{1/2}
\\
- \dfrac{T_{2}}{\zeta_{2}}\left( \dfrac{\zeta_{12}\zeta_{23}}{T_{12}T_{23}}\right)^{1/2} & 1
\end{pmatrix},
\label{Bmatrixthree}
\end{align}
respectively.
Then, according to Eq.~(\ref{Dmatrix}), the $2\times 2$ diffusion matrix ${\bf D}$ becomes 
\begin{align}
{\bf D}  = \begin{pmatrix} 
T_{12}/\zeta_{12} &  -T_2/\zeta_2
\\
 -T_2/\zeta_2 & T_{23}/\zeta_{23}
\end{pmatrix}.
\label{Dmatrixthree}
\end{align}

In our previous work, we obtained the steady-state probability distribution function 
$p_0({\bm r})$ for a three-sphere micromachine~\cite{Sou19}.
Owing to the reproductive property of Gaussian distributions~\cite{KampenBook,RiskenBook}, 
$p_{0}({\bm r})$ should also be a Gaussian function for the present linear problem and is given by 
\begin{align}
p_0({\bm r}) = {\cal N}_0 \exp \left[ -  \frac{1}{2} {\bm r}^{\mathsf T} {\mathbf C}_0^{-1} 
{\bm r} \right].
\label{gaussian}
\end{align}
In the above, ${\cal N}_0$ is the normalization factor, and the steady-state covariance matrix 
${\mathbf C}_0$ is given by
\begin{align}
{\mathbf C}_0 & = \frac{1}{K}
\begin{pmatrix} 
T_{12} + \zeta_{12}\Delta  & \zeta_2  \Delta  
\\
\zeta_2  \Delta  & T_{23} + \zeta_{23}\Delta
\end{pmatrix},
\label{covarianceC0}
\end{align}
with
\begin{align}
\Delta = \frac{\zeta_{12} \zeta_{23}(T_{12} + T_{23}-2T_2)}
{(\zeta_{12} + \zeta_{23})(\zeta_2^2 -\zeta_{12} \zeta_{23})}.
\end{align}

In the following sections, the above matrices are used to calculate the average irreversibility and 
the average entropy production rate.

\section{Irreversibility}
\label{sec:irreversibility}

\begin{figure}[tbh]
\begin{center}
\includegraphics[scale=0.28]{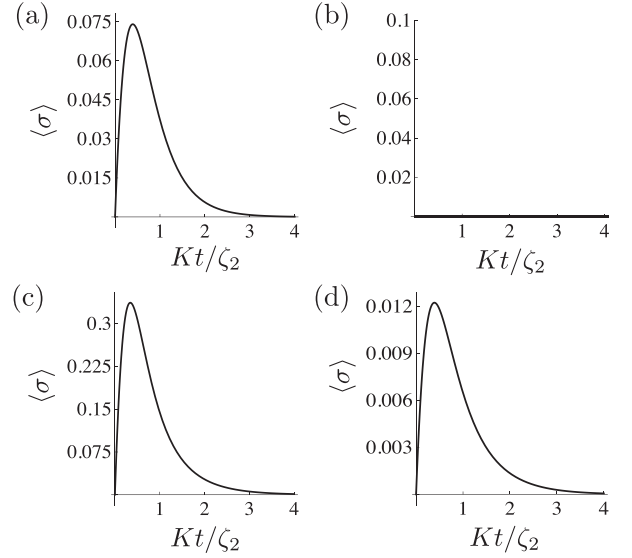}
\end{center}
\caption{The average irreversibility $\langle \sigma(t) \rangle$ given by Eq.~(\ref{meanirr}) as 
a function of dimensionless time $Kt/\zeta_{2}$.
The parameters are 
(a) $\zeta_1=\zeta_2 = \zeta_3$,
$\tau_1 = 1/900$, $\tau_2 = 41/900$, and  $\tau_3 = 81/900$;
(b) $\zeta_1=\zeta_2 = \zeta_3$,
$\tau_1 = \tau_3 = 25/900$, and  $\tau_2 = 41/900$;
(c) $\zeta_1/\zeta_2 = 0.5$, $\zeta_3/\zeta_2 = 5$,
$\tau_1 = 1/900$, $\tau_2 = 41/900$, and  $\tau_3 = 81/900$
(these temperatures are the same as in (a));
(d) $\zeta_1/\zeta_2 = 0.5$, $\zeta_3/\zeta_2 = 5$,
$\tau_1 = \tau_3 = 25/900$, and  $\tau_2 = 41/900$
(these temperatures are the same as in (b)).
}
\label{plotirr}
\end{figure}

In this section, we calculate the average  irreversibility $\langle \sigma (t) \rangle$ in 
Eq.~(\ref{meanirr}) for a thermally driven three-sphere micromachine by using 
${\bf A}$ and ${\bf C}_{0}$ in Eqs.~(\ref{Amatrixthree}) and (\ref{covarianceC0}), respectively
(notice that ${\bf C}_{t}$ is also given by ${\bf A}$ and ${\bf C}_{0}$ according to 
Eq.~(\ref{Cta})).
In Fig.~\ref{plotirr}, we plot $\langle \sigma (t) \rangle$ as a function of dimensionless time $K t/ \zeta_2$.
We also define the dimensionless temperature of the three spheres by 
\begin{align}
\tau_i = \frac{2T_i}{K\ell^2},
\end{align}
which is the ratio between the thermal energy of each sphere and the spring elastic energy
(recall $k_{\rm B}=1$).

\begin{figure}[tbh]
\begin{center}
\includegraphics[scale=0.6]{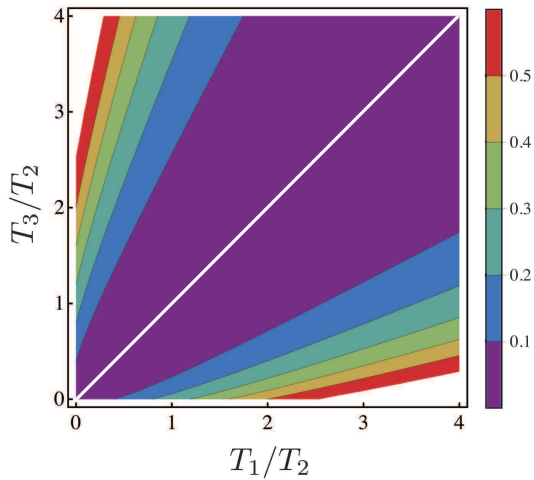}
\end{center}
\caption{(Color online)　The dimensionless average entropy production rate 
$\zeta \langle \dot{\sigma} \rangle / K$ given by Eq.~(\ref{entpro2}) as a function of 
$T_{1}/T_{2}$ and $T_{3}/T_{2}$.
The average entropy production rate vanishes along the diagonal white line $T_1=T_3$.}
\label{entropycontour}
\end{figure}

\begin{figure}[tbh]
\begin{center}
\includegraphics[scale=0.6]{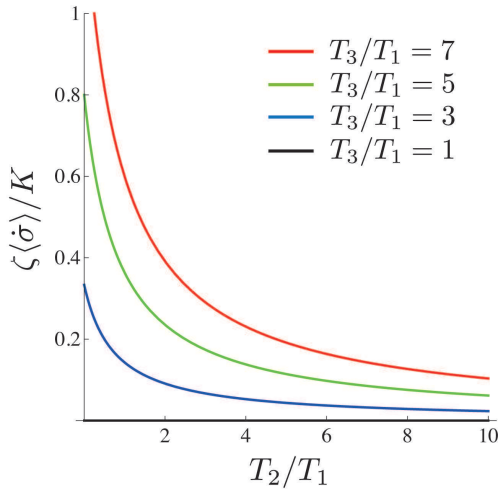}
\end{center}
\caption{(Color online)　The dimensionless entropy production rate $\zeta \langle \dot{\sigma} \rangle / K$ 
given by Eq.~(\ref{entpro2}) as a function of $T_{2}/T_{1}$ for  
$T_{3}/T_{1}=1, 3, 5$, and $7$.
}
\label{entropylines}
\end{figure}

\begin{figure}[tbh]
\begin{center}
\includegraphics[scale=0.6]{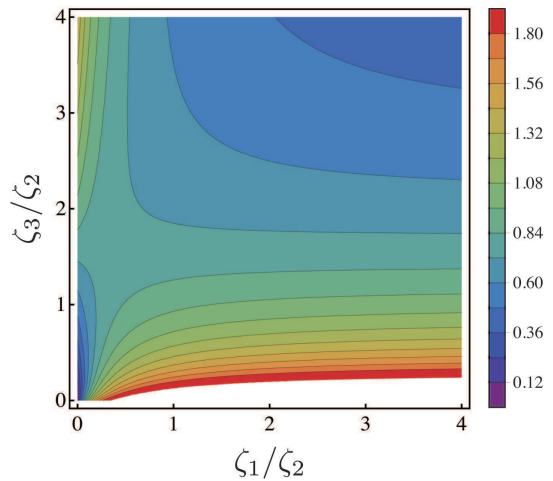}
\end{center}
\caption{(Color online)　The dimensionless entropy production rate $\zeta_2 \langle \dot{\sigma} \rangle / K$ 
given by Eq.~(\ref{entpro}) as a function of $\zeta_{1}/\zeta_{2}$ and $\zeta_{3}/\zeta_{2}$.
Here the temperature ratios are fixed to $T_{1}/T_{2} = 0.2$ and $T_{3}/T_{2} = 5$.
}
\label{entropyasym}
\end{figure}

The chosen parameters in Fig.~\ref{plotirr}(a) are 
$\zeta_1=\zeta_2 = \zeta_3$, 
$\tau_1 = 1/900$, $\tau_2 = 41/900$, and  $\tau_3 = 81/900$, and 
those in Fig.~\ref{plotirr}(b) are 
$\zeta_1= \zeta_2 = \zeta_3$,
$\tau_1 = \tau_3 = 25/900$, and  $\tau_2 = 41/900$. 
For these two cases, the friction coefficients are all identical. 
In Fig.~\ref{plotirr}(a) for which $\tau_1 \neq \tau_3$, the average irreversibility first increases 
from zero and it vanishes in the long time limit $t \rightarrow \infty$. 
This is because $p({\bm x}_{0},{\bm x}_{0},0) = p({\bm x}_{1},{\bm x}_{1},0)$ for $t=0$, whereas 
the transition probability becomes a stationary one for $t \rightarrow \infty$.
Since the average irreversibility is positive semidefinite for all time and goes to zero as $t \rightarrow 0$
and $t \rightarrow \infty$, there is a typical time $K t^{\ast}/ \zeta_2 \approx 0.39$
for which $\langle \sigma (t) \rangle $ takes a global maximum value. 
Notice that $\zeta_2/K$ corresponds to the spring relaxation time, and $t^{\ast}$ gives a 
characteristic time scale of the irreversible fluctuations.
When $\tau_1 = \tau_3$ as in Fig.~\ref{plotirr}(b), on the other hand, $\langle \sigma (t) \rangle$ 
vanishes for all $t$.
In this case, the system is in apparent equilibrium because the micromachine is thermally balanced~\cite{Sou19}.

Keeping the temperature parameters $\tau_i$ the same as in Figs.~\ref{plotirr}(a) and (b), 
we introduce asymmetry in the friction coefficients in Figs.~\ref{plotirr}(c) and (d) for which
we set $\zeta_1/\zeta_2 = 0.5$ and $\zeta_3/\zeta_2 = 5$.
Although the time evolutions in Fig.~\ref{plotirr}(a) and (c) are similar, the maximum value of 
$\langle \sigma \rangle$ in Fig.~\ref{plotirr}(c) is about $4.5$ times larger than that in 
Fig.~\ref{plotirr}(a).
Hence the asymmetry in the friction coefficients increases the average irreversibility.
In Fig.~\ref{plotirr}(d), the irreversibility is non-zero and $\langle \sigma(t) \rangle \ge 0$ even 
$\tau_1 = \tau_3$.
This is because the system is not thermally balanced, namely, $T_{12} \neq T_{23}$, and the 
micromachine is in out-of-equilibrium.
It is worth mentioning that the average irreversibility $\langle \sigma (t) \rangle$ depends only 
on the ratios of the temperatures such as $T_1/T_2$ and $T_3/T_2$.

In general, the irreversibility $\sigma$ is a measure of non-equilibrium fluctuations, and one can 
distinguish forward from reverse trajectories segments when $\sigma$ is non-zero.
The result in Fig.~\ref{plotirr} clearly indicates that the asymmetry of either the temperatures 
or the friction coefficients are necessary to drive a three-sphere micromachine out of equilibrium.
The maximum value of $\langle \sigma \rangle$ in Fig.~\ref{plotirr} directly reflects the 
magnitude of the irreversibility. 
It is useful to note that the asymmetry of both the temperatures and the friction coefficients
can enhance the maximum value of $\langle \sigma \rangle$ (compare Figs.~\ref{plotirr}(a) 
and (c)).
As we will discuss in the next section, the larger the maximum value of $\langle \sigma \rangle$ 
is, the larger the average entropy production rate of a thermally driven micromachine becomes.

\section{Entropy production rate}
\label{sec:entropy}

Next we calculate the steady-state average entropy production rate of a thermally driven micromachine. 
In the steady-state, the entropy production rate balances with the entropy extraction rate, and 
both quantities become zero at equilibrium~\cite{Taye15}.
Substituting ${\bf A}$, ${\bf D}$, and ${\bf C}_{0}$ in Eqs.~(\ref{Amatrixthree}), (\ref{Dmatrixthree}), 
and (\ref{covarianceC0}), respectively, to Eqs.~(\ref{entproorig}) and (\ref{gain}), we obtain after 
some calculation 
\begin{align}
\langle \dot{\sigma} \rangle = \frac{K  [\zeta_{1}(T_{3}-T_{2}) + \zeta_{3}(T_{2} - T_{1})]^{2}}
{(\zeta_{1}T_{2}T_{3} + \zeta_{2}T_{3}T_{1}+\zeta_{3}T_{1}T_{2})
(\zeta_{1}\zeta_{2} + \zeta_{2}\zeta_{3} + 2\zeta_{1}\zeta_{3})}.
\label{entpro}
\end{align}
This is an important result of this paper. 
Obviously, we have $\langle \dot{\sigma} \rangle \ge 0$.
When the system is in thermal equilibrium, i.e., $T_{1}=T_{2}=T_{3}$, the entropy production
rate vanishes for any combination of the friction coefficients.
We note that Eq.~(\ref{entpro}) does not depend on the spring natural length $\ell$ because 
we are considering only the fluctuations around $\ell$ as defined in Eq.~(\ref{r12r23}).

When the three friction coefficients are all identical, i.e., $\zeta_{1} = \zeta_{2} = \zeta_{3}=\zeta$,
Eq.~(\ref{entpro}) reduces to 
\begin{align}
\langle \dot{\sigma} \rangle = \frac{K (T_{1} - T_{3})^{2}}{4 \zeta (T_{1}T_{2} + T_{2}T_{3} + T_{3}T_{1})}
\label{entpro2}.
\end{align}
Using Eq.~(\ref{entpro2}), we give in Fig.~\ref{entropycontour} a color representation of the 
dimensionless average entropy production rate $\zeta \langle \dot{\sigma} \rangle/ K$ as 
a function of $T_1/T_2$ and $T_3/T_2$.
Here $\langle \dot{\sigma} \rangle$ vanishes when $T_{1}=T_{3}$ and it increases as the 
difference between $T_{1}$ and $T_{3}$ becomes larger.
To see the role of the temperature $T_{2}$ of the middle sphere in Eq.~(\ref{entpro2}), we plot in 
Fig.~\ref{entropylines} the steady-state average entropy production rate
$\zeta \langle \dot{\sigma} \rangle/ K$ as a function of  $T_{2}/T_{1}$ (not $T_1/T_2$)
for four different values of $T_{3}/T_{1}$.
From this plot, one can clearly see that $\langle \dot{\sigma} \rangle$ becomes smaller as $T_{2}/T_{1}$ 
is increased.

When the friction coefficients are different between the three spheres, Eq.~(\ref{entpro}) implies that 
the average entropy production rate is non-zero, $\langle \dot{\sigma} \rangle>0$, even when 
$T_{1}=T_{3}$.
In general, $\langle \dot{\sigma} \rangle$ becomes larger when  
either $\zeta_{1}/\zeta_{2}$ or $\zeta_{3}/\zeta_{2}$ is increased.
As an example of asymmetric situations, we give in Fig.~\ref{entropyasym} a color representation  
of the dimensionless average entropy production rate $\zeta_2 \langle \dot{\sigma} \rangle/ K$ 
as a function of $\zeta_{1}/\zeta_{2}$ and $\zeta_{3}/\zeta_{2}$ when 
$T_{1} / T_{2} = 0.2$ and $T_{3} / T_{2} = 5$.
The value of $\langle \dot{\sigma} \rangle$ is asymmetric with respect to the line 
$\zeta_{1}= \zeta_{3}$, and it becomes larger when $\zeta_{3}/\zeta_{2}$ becomes smaller.

In multidimensional systems, it was generally discussed that enhanced fluctuation
(or noise amplification) occurs not through some additional forcing, but through 
violation of detailed balance which is measured by the gain matrix ${\bf G}$ in Eq.~(\ref{gain})~\cite{Weiss03}.
Hence, according to Eq.~(\ref{entproorig}), the entropy production rate reflects the noise amplification that occurs when detailed balance is not satisfied~\cite{Weiss07}.
It is also known that the entropy production rate is related to the heat flow in 
a system~\cite{Li19}.

As we can see from Eq.~(\ref{Vthermal}) for the average velocity $\langle V \rangle$ and 
Eq.~(\ref{entpro2}) for the average entropy production rate $\langle \dot{\sigma} \rangle$
of a thermally driven three-sphere micromachine, the temperature difference between the first 
and the third spheres, $T_1-T_3$, plays an essential role to characterize its non-equilibrium 
behaviors.
According to their dependence on $T_1-T_3$, we see a proportionality relation such that
$\langle \dot{\sigma} \rangle \sim \langle V \rangle^2$.
In our previous work~\cite{Hosaka17}, we showed that the average velocity $\langle V \rangle$ 
of a three-sphere micromachine is proportional to the net heat flow between the first and 
the third spheres.

\section{Diffusion coefficient}
\label{sec:diffusion}

In this section, we discuss the Brownian motion of a thermally driven three-sphere micromachine. 
We introduce the center of \textit{mass} position of a micromachine by 
\begin{align}
X(t) = \frac{1}{3}[x_{1}(t) + x_{2}(t) + x_{3}(t)],
\label{com}
\end{align}
even though we neglect the inertia of the spheres.
From Eqs.~(\ref{x1dot})--(\ref{x3dot}), the Langevin equation for $X$ can be written in terms of 
$r_{12}$ and $r_{23}$ as 
\begin{align}
\frac{d X}{dt} & =  \frac{K}{3}\left( \frac{\zeta_{2} - \zeta_{1}}{\zeta_{1}\zeta_{2}} \right) r_{12} 
+ \frac{K}{3}\left( \frac{\zeta_{3} - \zeta_{2}}{\zeta_{2}\zeta_{3}} \right) r_{23}
\nonumber \\
& + \frac{\sqrt{2}}{3}\left( \frac{T_{1}}{\zeta_{1}} + \frac{T_{2}}{\zeta_{2}} + \frac{T_{3}}{\zeta_{3}} \right)^{1/2} 
\xi_{X},
\label{efflangevin}
\end{align}
where $\xi_{X}(t)$ is a zero mean and unit variance Gaussian white noise defined by 
\begin{align}
\xi_{X}&  = \left( \frac{T_{1}}{\zeta_{1}} + \frac{T_{2}}{\zeta_{2}} + \frac{T_{3}}{\zeta_{3}} \right)^{{-1/2}}
\nonumber \\
& \times \left[ 
\left( \frac{T_1}{\zeta_1} \right)^{{1/2} }\xi_{1} + 
\left( \frac{T_2}{\zeta_2} \right)^{{1/2} }\xi_{2} +
\left( \frac{T_3}{\zeta_3} \right)^{{1/2} }\xi_{3} \right], 
\label{timex}
\end{align}
and satisfies the following statistical properties
\begin{align}
& \langle \xi_{X}(t) \rangle  = 0,
\\
& \langle \xi_{X}(t) \xi_{X}(t') \rangle = \delta (t-t').
\end{align}

Using Eqs.~(\ref{eqr12}) and (\ref{eqr23}) for the dynamics of $r_{12}$ and $r_{23}$, respectively, the 
mean squared displacement of the center of mass position becomes 
\begin{align}
\langle X^2(t)\rangle = 2 D t, 
\end{align}
where the diffusion coefficient is obtained as 
\begin{align}
D  = \frac{\zeta_{1} T_{1} + \zeta_{2} T_{2} + \zeta_{3} T_{3}}{(\zeta_{1} + \zeta_{2} + \zeta_{3})^{2}}.
\label{diffsion3}
\end{align}
See Appendix~\ref{app:msd} for the detailed derivation.
When the temperatures are all identical, $T_{1}=T_{2}=T_{3}$, Eq.~(\ref{diffsion3}) becomes
\begin{align}
D  = \frac{T_{2}}{\zeta_{1} + \zeta_{2} + \zeta_{3}},
\label{diffsionequilibrium}
\end{align}
as expected for the equilibrium case.

On the other hand, when the three friction coefficients are all identical, 
$\zeta_{1} = \zeta_{2} = \zeta_{3}=\zeta$, Eq.~(\ref{diffsion3}) reduces to 
\begin{align}
D  = \frac{T_{1} + T_{2} + T_{3}}{9\zeta},
\label{diffsion3sym}
\end{align}
which is proportional to the sum of the three temperatures. 
Introducing an average temperature by 
\begin{align}
T_X = \frac{T_{1} + T_{2} + T_{3}}{3},
\label{averagetemp}
\end{align}
we can rewrite Eq.~(\ref{diffsion3sym}) as $D  = T_X/(3\zeta)$.

\begin{figure}[bth]
\begin{center}
\includegraphics[scale=0.6]{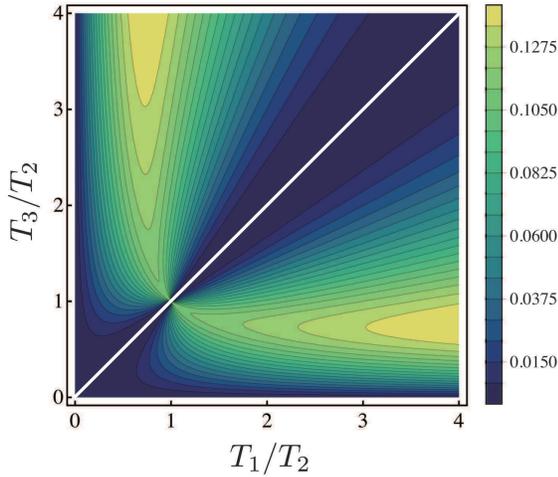}
\end{center}
\caption{(Color online)　The scaled efficiency $128 \ell^2 \varepsilon/(27 a^2 \tau_2)$ 
given by Eq.~(\ref{efficiencyresult}) as a function of $T_{1}/T_{2}$ and $T_{3}/T_{2}$.
The efficiency vanishes along the diagonal white line $T_1=T_3$.}
\label{efficiencyplot}
\end{figure}

In general, the diffusion coefficient of a thermally driven $n$-sphere swimmer is predicted to be 
\begin{align}
D_{n} = \frac{\sum_{i = 1}^{n} \zeta_{i}T_{i}}{\left( \sum_{i=1}^{n} \zeta_{i} \right)^{2}}.
\label{nspherediff}
\end{align}
This expression can be explicitly confirmed also for $n=2$.
For $n=2$, the diffusion coefficient for the center of \textit{friction} was obtained in Ref.~\cite{Grosberg15},
and it is different from that of the center of \textit{mass} diffusion.
This difference is not physically essential because it only depends on the choice of the 
coordinate system.
Nevertheless, we consider that the center of mass diffusion is much easier to be measured in 
the experiments.

\section{Efficiency}
\label{sec:efficiency}

Finally, we shall estimate the efficiency of a thermally driven micromachine. 
When the three friction coefficients are all identical and given $\zeta$, the average velocity
is given by Eq.~(\ref{Vthermal}) which can be rewritten as~\cite{Hosaka17} 
\begin{align}
\langle V\rangle=\frac{k_{\rm B}(T_3-T_1)a}{16 \zeta \ell^2}.
\label{Vsymmetric}
\end{align}
Here $a$ is the radius of the spheres, and we employ the Stokes relation $\zeta = 6 \pi \eta a$
for the friction coefficient.
Notice that we recover the Boltzmann constant $k_{\rm B}$ in this section for the sake of clarity.
The above result indicates that the swimming direction is from a colder sphere to a hotter one, and 
the velocity does not depend on the temperature of the middle sphere~\cite{Hosaka17}.

Following Ref.~\cite{Golestanian08}, we define the efficiency of a thermally driven 
micromachine by
\begin{align}
\varepsilon = \frac{3\zeta \langle V\rangle^2}{k_{\rm B} \langle \dot{\Sigma} \rangle T_X},
\label{efficiencydef}
\end{align}
where $\langle \dot{\Sigma} \rangle$ is the total entropy production rate and 
the average temperature $T_X$ is given by Eq.~(\ref{averagetemp}).
Notice that the entropy production rate $\langle \dot{\sigma} \rangle$ in Eq.~(\ref{entpro2}) 
takes into account only the internal motions ($r_{12}$ and $r_{23}$) of a micromachine
and it also vanishes when $T_1=T_3$.

In order to obtain $\langle \dot{\Sigma} \rangle$, one needs to take into account 
the center of mass motion $X$ given by Eq.~(\ref{com}), and solve Eqs.~(\ref{eqr12}),
(\ref{eqr23}), and (\ref{efflangevin}) simultaneously. 
For the three-dimensional vectors ${\bm r} = (r_{12}, r_{23}, X)^{\mathsf T}$ and 
${\bm \xi} = (\xi_{12} , \xi_{23}, \xi_X)^{\mathsf T}$, 
the corresponding $3\times 3$ matrices ${\bf A}$, ${\bf F}$, and ${\bf D}$ are shown in 
Appendix~\ref{app:3dimmatrix}.
Repeating the same calculation as in Sec.~\ref{sec:entropy}, we obtain the following total entropy 
production rate 
\begin{align}
\langle \dot{\Sigma} \rangle & = \frac{K}{12 \zeta T_{1} T_{2} T_{3}}
(T_{1}^{2} T_{2} + 4 T_{1} T_{2}^{2} + 4 T_{1}^{2} T_{3} 
\nonumber \\
& + T_{3}^{2}T_{2} + 4 T_{3}T_{2}^{2} + 4 T_{1} T_{3}^{2} - 18 T_{1} T_{2} T_{3}),
\label{totalentropy}
\end{align} 
when the friction coefficients are identical. 
Unlike $\langle \dot{\sigma} \rangle$ in Eq.~(\ref{entpro2}), $\langle \dot{\Sigma} \rangle$
in Eq.~(\ref{totalentropy}) vanishes only when $T_1=T_2=T_3$, i.e., thermal equilibrium.
It should be noted here that, although the above total entropy production rate 
$\langle \dot{\Sigma} \rangle$ takes into account all the three positional degrees of freedom 
of the spheres, it still does not include hydrodynamic interactions acting between different 
spheres, which have been neglected throughout this paper.
Hence, the maximum of $\varepsilon$ in Eq.~(\ref{efficiencydef}) is not necessarily unity. 
Nevertheless, $\varepsilon$ gives a useful measure of the ratio between the hydrodynamic 
dissipation of a micromachine and its total entropy production rate due to the sphere motions.

With the above result, the efficiency $\varepsilon$ in Eq.~(\ref{efficiencydef}) can be obtained as 
\begin{widetext}
\begin{align}
\varepsilon = \frac{27 a^{2} k_{\rm{B}} T_{1} T_{2}T_{3}(T_{1}-T_{3})^{2}}
{64 K \ell^{4}(T_{1} + T_{2} + T_{3})(T_{1}^{2} T_{2} + 4 T_{1} T_{2}^{2} + 4 T_{1}^{2} T_{3}
+ T_{3}^{2}T_{2} + 4 T_{3}T_{2}^{2} + 4 T_{1} T_{3}^{2} - 18 T_{1} T_{2} T_{3})}.
\label{efficiencyresult}
\end{align}
\end{widetext}
Clearly, $\varepsilon$ vanishes when $T_1=T_3$ as it should.
The important outcome of Eq.~(\ref{efficiencyresult}) is that the efficiency scales as 
$(a/\ell)^2$ and is proportional to the temperature $T_2$ of the middle sphere.
By using the dimensionless temperature $\tau_2 = 2T_2/(K\ell^2)$, we give in 
Fig.~\ref{efficiencyplot} a color representation of the scaled efficiency 
$128 \ell^2 \varepsilon/(27 a^2 \tau_2)$ as a function of $T_1/T_2$ and $T_3/T_2$.
The efficiency also becomes smaller when the temperatures are highly asymmetric.

On the other hand, the efficiency $\varepsilon$ becomes larger along a certain characteristic 
curve.
Such a dependence on the temperatures can be explained as follows.
The numerator of $\varepsilon$ (hydrodynamic dissipation) always vanishes for 
$T_1=T_3$, and all the contour lines are parallel to the line $T_1=T_3$.
On the other hand, the denominator of $\varepsilon$ (total entropy production rate) 
increases when $T_1$ and $T_3$ are either small or large even for $T_1=T_3$
(the minimum occurs at $T_1=T_2=T_3$), and the contour lines are perpendicular to the 
line $T_1=T_3$.
These different temperature dependences between the numerator and the denominator 
give rise to the nontrivial increase of $\varepsilon$ as shown in Fig.~\ref{efficiencyplot}.

\section{Summary and discussion}
\label{sec:discussion}

In this paper, we have discussed the non-equilibrium behaviors of a thermally driven 
elastic three-sphere micromachine.
In our model, the three spheres are in contact with independent heat baths having 
different temperatures~\cite{Hosaka17,Sou19}.
Using the formulation of a linear stochastic Langevin dynamics~\cite{Weiss03,Weiss07}, 
we have calculated the time-dependent average irreversibility $\langle \sigma (t) \rangle$
in Eq.~(\ref{meanirr}).
When the temperatures and/or the friction coefficients are asymmetric, the average irreversibility
is non-zero and takes a global maximum value for a finite time.
The corresponding characteristic time scale is roughly set by the spring relaxation time $\zeta_{2}/K$.
We have further obtained the average entropy production rate $\langle \dot{\sigma} \rangle$ 
in Eq.~(\ref{entpro}) which is the zero-time growth rate of the average irreversibility.
This quantity decreases as the temperature of the middle phase increases.

We have also discussed the Brownian motion of a thermally driven three-sphere micromachine 
and calculated its center of \textit{mass} diffusion coefficient $D$ as in Eq.~(\ref{diffsion3}).
The obtained expression can be generalized for a many-sphere micromachine.
When the friction coefficients are identical, an average temperature $T_X$ can be introduced
as in Eq.~(\ref{averagetemp}).
Our result is different from the diffusion coefficient for the center of \textit{friction}  
obtained for a non-equilibrium dimer model~\cite{Grosberg15}. 
Finally, with the results of the total entropy production rate and the average temperature, 
we have estimated the efficiency of a micromachine in Eq.~(\ref{efficiencyresult}).

Our model of a three-sphere micromachine has a similarity to that of two over-damped, 
tethered spheres coupled by a harmonic spring and also confined between two 
walls~\cite{Battle16,Gnesotto18}. 
In these works, the authors numerically showed that displacements obey a Gaussian distribution 
and also found probability flux loops that demonstrate the broken detailed balance~\cite{Battle16,Gnesotto18}.
The two displacements $r_{12}$ and $r_{23}$ in Eq.~(\ref{r12r23}) correspond 
to the sphere positions in their model.
However, the presence of the middle sphere changes the structure of the frequency matrix
for a three-sphere micromachine when $T_2 \neq 0$~\cite{Sou19}.
Moreover, a two-sphere micromachine in a viscous fluid cannot have a directed motion 
even if the temperatures are different~\cite{Hosaka17}.
Recently, Li \textit{et al.}\ used the two-sphere model to calculate the entropy 
production rate~\cite{Li19}.
We note that our result in Eq.~(\ref{entpro2}) reduces to their expression when $T_2=0$.

In this work, we have neglected long-ranged hydrodynamic interactions acting between different 
spheres and we have not considered the locomotion of a micomachine~\cite{Hosaka17}.
If hydrodynamic interactions are taken into account in the present analysis, the covariance 
matrix in Eq.~(\ref{covarianceC0}) is modified in non-equilibrium situations.
Such hydrodynamic corrections should be proportional to $a/\ell$ within the lowest-order expansion.
Moreover, these corrections should vanish in thermal equilibrium, i.e., $T_1=T_2=T_3$ because 
hydrodynamic interactions should not affect equilibrium statistical properties.

\acknowledgements

Y.H.\ acknowledges support by a Grant-in-Aid for JSPS Fellows (Grant No.\ 19J20271) from the Japan Society 
for the Promotion of Science (JSPS). 
K.Y.\ acknowledges support by a Grant-in-Aid for JSPS Fellows (Grant No.\ 18J21231) from the JSPS.
S.K.\ acknowledges support by a Grant-in-Aid for Scientific Research (C) (Grant No.\ 18K03567 and
Grant No.\ 19K03765) from the JSPS, 
and support by a Grant-in-Aid for Scientific Research on Innovative Areas
``Information physics of living matters'' (Grant No.\ 20H05538) from the Ministry of Education, Culture, 
Sports, Science and Technology of Japan.

\begin{widetext}
\appendix
\section{Irreversibility distribution $P(\sigma)$}
\label{app:irrpdf}

Following Ref.~\cite{Weiss07}, we show how to calculate the probability density function of irreversibility 
$P(\sigma)$ for a system described by the linear Langevin dynamics in Eq.~(\ref{langevingeneral}). 
We remind that $P(\sigma)$ satisfies the fluctuation theorem in Eq.~(\ref{fluctuation}).

We first introduce a $2N$-dimensional state space  
\begin{align}
{\bm z} & = 
\begin{pmatrix} 
{\bm r}_0
\\
{\bm r}_1
\end{pmatrix}, 
\label{zvectora}
\end{align}
where ${\bm r}_0$ and ${\bm r}_1$ are the initial and final states.
The probability density function of irreversibility $P(\sigma)$ is given by 
\begin{align}
P(\sigma) = \int d^{2N}{\bm z} \, \delta(\sigma - {\bm z}^{\mathsf T} {\bf R} {\bm z}/2) 
p({\bm r}_{0}, {\bm r}_{1},t),
\label{pra}
\end{align}
where ${\bf R} = {\bf R}_{10} - {\bf R}_{01}$ is the $2N \times 2N$ matrix and we have 
used the matrices
\begin{align}
{\bm R}_{10} & = 
\begin{pmatrix} 
{\bf C}_{t}^{-1} & -{\bf C}_{t}^{-1} e^{{\bf A}t}
\\
-e^{{\bf A}^{\mathsf{T}}t} {\bf C}_{t}^{-1} 
& e^{{\bf A}^{\mathsf{T}}t}{\bf C}_{t}^{-1} e^{{\bf A} t} + {\bf C}_{0}^{-1} 
\end{pmatrix}, 
\label{r10matrixa}
\end{align}
and 
\begin{align}
{\bm R}_{01} & = 
\begin{pmatrix} 
e^{{\bf A}^{\mathsf{T}}t}{\bf C}_{t}^{-1} e^{{\bf A} t} + {\bf C}_{0}^{-1}  
& -e^{{\bf A}^{\mathsf{T}}t} {\bf C}_{t}^{-1}
\\
- {\bf C}_{t}^{-1} e^{{\bf A}t} & {\bf C}_{t}^{-1}
\end{pmatrix}.
\label{r01matrixa}
\end{align}

The characteristic function of the probability density function is defined by 
\begin{align}
P[k] = \int d{\rm \sigma} \, P(\sigma) e^{ik\sigma}.
\label{fourier}
\end{align}
For linear Langevin systems, it is shown that the characteristic function can be expressed 
as~\cite{Weiss07} 
\begin{align}
P[k] =  \frac{1}{\prod_{m = 1}^{2N} \sqrt{1 - i k \lambda_m}},
\label{prklasta}
\end{align}
where $\lambda_m$ are the $2N$ eigenvalues of the matrix ${\bm R}_{01}^{-1}{\bm R}$.
Then the probability density function of irreversibility in Eq.~(\ref{pra}) can be obtained by 
the inverse transform of Eq.~(\ref{invfourier}):  
\begin{align}
P(\sigma) = \frac{1}{2\pi} \int dk \, P[k] e^{-ik\sigma}.
\label{invfourier}
\end{align}

We discuss here the four eigenvalues for a thermally driven micromachine. 
When $T_1=T_3$, we find that all the eigenvalues vanish, i.e., $\lambda_m=0$.
In this thermally balanced situation, the characteristic function is simply $P[k]=1$ and 
the probability density function of irreversibility is $P(\sigma)=\delta(\sigma)$.

As the simplest non-equilibrium situation, we consider the case when $T_1=T_2=0$ but 
$T_3 \neq 0$. 
Then the four eigenvalues can be obtained in the short time limit as 
\begin{align}
\lambda_{1} &\approx \frac{\sqrt{3}}{8}(2 \bar{t}-1) (\bar{t}+2) \sqrt{\bar{t} (23 \bar{t}+8)},
\\
\lambda_{2} &\approx -\frac{\sqrt{3}}{8} (2 \bar{t}-1) (\bar{t}+2) \sqrt{\bar{t} (23 \bar{t}+8)},
\\
\lambda_{3} &\approx -\frac{(2\bar{t}-1)}{\bar{t}}\biggl{[} 3 - \frac{\sqrt{3}}{8} (\bar{t}+2)^{2} \sqrt{(\bar{t}+2) (7\bar{t} +6)} \biggl{]},
\\
\lambda_{4} &\approx -\frac{(2\bar{t}-1)}{\bar{t}}\biggl{[} 3 + \frac{\sqrt{3}}{8} (\bar{t}+2)^{2} \sqrt{(\bar{t}+2) (7\bar{t} +6)} \biggl{]},
\end{align}
where $\bar{t}=Kt/\zeta_2$ is the dimensionless time.
The above expressions are valid when $\bar{t} \ll 1$.

\section{Derivation of diffusion coefficient}
\label{app:msd}

In this Appendix, we show the derivation of the diffusion coefficient in Eq.~(\ref{diffsion3}). 
We first integrate Eq.~(\ref{efflangevin}) over time, and obtain the mean squared displacement of a 
three-sphere micromachine as
\begin{align}
\langle X(t)^{2} \rangle & = \frac{K^{2}}{9} \biggl{(} \frac{\zeta_{2} - \zeta_{1}}{\zeta_{1}\zeta_{2}} \biggl{)} ^{2} \int_{0}^{t} dt_{1} \int_{0}^{t} dt_{2} \, \langle r_{12}(t_{1}) r_{12}(t_{2}) \rangle 
\nonumber \\
& + \frac{K^{2}}{9} \biggl{(} \frac{\zeta_{3} - \zeta_{2}}{\zeta_{2}\zeta_{3}} \biggl{)} ^{2} \int_{0}^{t} dt_{1} \int_{0}^{t} dt_{2} \, \langle r_{23}(t_{1}) r_{23}(t_{2}) \rangle 
\nonumber \\
& + \frac{2}{9}\biggl{(}  \frac{T_{1}}{\zeta_{1}} + \frac{T_{2}}{\zeta_{2}} + \frac{T_{3}}{\zeta_{3}} \biggl{)} \int_{0}^{t} dt_{1} \int_{0}^{t} dt_{2} \, \langle \xi_{X}(t_{1}) \xi_{X} (t_{2}) \rangle 
\nonumber \\
& + \frac{2K^{2}}{9}  \biggl{(} \frac{\zeta_{2} - \zeta_{1}}{\zeta_{1}\zeta_{2}} \biggl{)} \biggl{(} \frac{\zeta_{3} - \zeta_{2}}{\zeta_{2}\zeta_{3}} \biggl{)}  \int_{0}^{t} dt_{1} \int_{0}^{t} dt_{2} \, \langle r_{12}(t_{1}) r_{23}(t_{2}) \rangle 
\nonumber \\
& + \frac{2\sqrt{2}K}{9} \biggl{(} \frac{\zeta_{2} - \zeta_{1}}{\zeta_{1}\zeta_{2}} \biggl{)}  
\biggl{(} \frac{T_{1}}{\zeta_{1}} + \frac{T_{2}}{\zeta_{2}} + \frac{T_{3}}{\zeta_{3}} \biggl{)}^{1/2} \int_{0}^{t} dt_{1} \int_{0}^{t} dt_{2} \, \langle r_{12}(t_{1}) \xi_{X}(t_{2}) \rangle 
\nonumber \\
& + \frac{2\sqrt{2}K}{9} \biggl{(} \frac{\zeta_{3} - \zeta_{2}}{\zeta_{2}\zeta_{3}} \biggl{)}  
\biggl{(} \frac{T_{1}}{\zeta_{1}} + \frac{T_{2}}{\zeta_{2}} + \frac{T_{3}}{\zeta_{3}} \biggl{)}^{1/2} \int_{0}^{t} dt_{1} \int_{0}^{t} dt_{2} \, \langle r_{23}(t_{1}) \xi_{X}(t_{2}) \rangle. 
\label{Xb}
\end{align}
Our task is to calculate the various noise correlation functions in the above expression.

Let us introduce the Fourier transform of a function $f(t)$ by 
\begin{align}
f[\omega] = \int_{-\infty}^{\infty} dt \, f(t) e^{i \omega t},~~~~~
f(t) = \frac{1}{2\pi} \int_{-\infty}^{\infty} d\omega \, f[\omega] e^{-i \omega t}. 
\label{fourierb}
\end{align}
Then we can solve the Langevin equations in Eqs.~(\ref{eqr12}) and (\ref{eqr23}) in the Fourier domain as
\begin{align}
r_{12}[\omega] & = - \frac{\left( \dfrac{\zeta_{2}}{\zeta_{23}} + \dfrac{i \zeta_{2} \omega}{K} \right) 
\left( \dfrac{2 T_{12}}{\zeta_{12}} \right)^{1/2} 
\xi_{12}[\omega] + 
\left( \dfrac{2T_{23}}{\zeta_{23}} \right)^{{1/2}} 
\xi_{23}[\omega]}{\dfrac{\zeta_{2}}{K} \omega^{2} - i \zeta_{2}\left( \dfrac{1}{\zeta_{12}} + \dfrac{1}{\zeta_{23}} \right) \omega + \dfrac{K}{\zeta_{2}} \left( 1 - \dfrac{\zeta_{2}^{2}}{\zeta_{12} \zeta_{23}} \right)}, 
\\
r_{23}[\omega] & = - \dfrac{\left(\dfrac{\zeta_{2}}{\zeta_{12}} + \dfrac{i \zeta_{2} \omega}{K} \right) 
\left( \dfrac{2 T_{23}}{\zeta_{23}}\right)^{1/2} \xi_{23}[\omega] + 
\left( \dfrac{2T_{12}}{\zeta_{12}} \right)^{1/2}  \xi_{12}[\omega]}
{\dfrac{\zeta_{2}}{K} \omega^{2} - i \zeta_{2}\left(  \dfrac{1}{\zeta_{12}} + \dfrac{1}{\zeta_{23}} \right) \omega + \dfrac{K}{\zeta_{2}} \left( 1 - \dfrac{\zeta_{2}^{2}}{\zeta_{12} \zeta_{23}} \right)}.
\label{rfreqb}
\end{align}

Calculating the products of the noise and taking the average, we obtain for example
\begin{align}
\langle r_{12}(t_{1})r_{12}(t_{2}) \rangle & = \left[ \frac{(\zeta_{2} \zeta_{12} + \zeta_{2} \zeta_{23} + \zeta_{12}\zeta_{23}H) T_{12}}{2 K H \zeta_{12}(\zeta_{12} + \zeta_{23})} - \frac{2 \zeta_{12} (\zeta_{2}^{2} T_{12} + \zeta_{12}\zeta_{23}T_{23} - 2 \zeta_{12} \zeta_{23} T_{2})}{K H (\zeta_{12}+ \zeta_{23} )(\zeta_{2}\zeta_{12} + \zeta_{2} \zeta_{23} +   \zeta_{12} \zeta_{23} H)} \right]
\nonumber \\
& \times  \exp \left[ - \frac{K(\zeta_{2} \zeta_{12} + \zeta_{2}\zeta_{23} + \zeta_{12} \zeta_{23} H)}{2 \zeta_{2} \zeta_{12} \zeta_{23} } |t_{1} - t_{2}| \right]
\nonumber \\
& + \left[ - \frac{(\zeta_{2} \zeta_{12} + \zeta_{2} \zeta_{23} - \zeta_{12}\zeta_{23}H) T_{12}}{2 K H \zeta_{12}(\zeta_{12} + \zeta_{23})} +\frac{2 \zeta_{12} (\zeta_{2}^{2} T_{12} + \zeta_{12}\zeta_{23}T_{23} - 2 \zeta_{12} \zeta_{23} T_{2})}{K H (\zeta_{12}+ \zeta_{23} )(\zeta_{2}\zeta_{12} + \zeta_{2} \zeta_{23} -  \zeta_{12} \zeta_{23} H)} \right]
\nonumber \\
& \times  \exp \left[ - \frac{K(\zeta_{2} \zeta_{12} + \zeta_{2}\zeta_{23} - \zeta_{12} \zeta_{23} H)}{2 \zeta_{2} \zeta_{12} \zeta_{23} } |t_{1} - t_{2}| \right],
\end{align}
where 
\begin{align}
H = \left( 4 + \frac{\zeta_{2}^{2}}{\zeta_{12}^{2}} + \frac{\zeta_{2}^{2}}{\zeta_{23}^{2}} - \frac{2 \zeta_{2}^{2}}{\zeta_{12} \zeta_{23}} \right)^{1/2}.
\end{align}
The other noise correlation functions can be obtained in a similar way.
However, it should be noted that, for non-equilibrium situations, the time-reversal invariance is not 
generally satisfied for the cross correlation functions, i.e., 
$\langle r_{12}(t_{1}) r_{23}(t_{2}) \rangle \neq \langle r_{12}(t_{2}) r_{23}(t_{1}) \rangle$.
On the other hand, the time-translational invariance of the noise correlation functions is always 
satisfied because we are dealing with steady-states.

 Collecting all the noise correlation functions and taking the limit of $t \rightarrow \infty$, we 
 finally obtain 
\begin{align}
D= \lim_{t \rightarrow \infty}\frac{\langle X^{2}(t) \rangle}{2t}
& =  \frac{\zeta_{12} (\zeta_{1} - \zeta_{2})^{2} (\zeta_{2}^{2} T_{12} + \zeta_{12} \zeta_{23} T_{23} - 2 \zeta_{12} \zeta_{23} T_{2})}{9\zeta_{1}^{2} (\zeta_{2}^{2} - \zeta_{12} \zeta_{23})^{2}} 
\nonumber \\
& + \frac{\zeta_{23} (\zeta_{3} - \zeta_{2})^{2} (\zeta_{2}^{2} T_{23} + \zeta_{12} \zeta_{23} T_{12} - 2 \zeta_{12} \zeta_{23} T_{2})}{9\zeta_{3}^{2} (\zeta_{2}^{2} - \zeta_{12} \zeta_{23})^{2}} 
\nonumber \\
& + \frac{1}{9} \biggl{(} \frac{T_{1}}{\zeta_{1}} + \frac{T_{2}}{\zeta_{2}} + \frac{T_{3}}{\zeta_{3}} \biggl{)} 
\nonumber \\
& + \frac{2\zeta_{12} \zeta_{23} (\zeta_{1} - \zeta_{2})(\zeta_{2} - \zeta_{3}) (\zeta_{2}^{2} T_{12} + \zeta_{2}^{2} T_{23} - \zeta_{2}^{2} T_{2} - \zeta_{12} \zeta_{23} T_{2})}{9\zeta_{1} \zeta_{2} \zeta_{3} (\zeta_{2}^{2} - \zeta_{12} \zeta_{23})^{2}} 
\nonumber \\
& + \frac{2\zeta_{12} (\zeta_{2} - \zeta_{1}) (\zeta_{1} \zeta_{2} \zeta_{23} T_{3} - \zeta_{1} \zeta_{3} \zeta_{23} T_{2} + \zeta_{1} \zeta_{2} \zeta_{3} T_{2} - \zeta_{2}^{2} \zeta_{3} T_{1})}{9\zeta_{1}^{2} \zeta_{2} \zeta_{3} (\zeta_{2}^{2} - \zeta_{12} \zeta_{23})} 
\nonumber \\
& + \frac{2\zeta_{23} (\zeta_{2} - \zeta_{3}) (\zeta_{2} \zeta_{3} \zeta_{12} T_{1} - \zeta_{1} \zeta_{3} \zeta_{12} T_{2} + \zeta_{1} \zeta_{2} \zeta_{3} T_{2} - \zeta_{1} \zeta_{2}^{2} T_{3})}{9\zeta_{1} \zeta_{2} \zeta_{3}^{2} (\zeta_{2}^{2} - \zeta_{12} \zeta_{23})}.
\label{difftrms}
\end{align}
Notice that the different lines in Eq.~(\ref{Xb}) correspond to the different lines in the above equation.
The right hand side of Eq.~(\ref{difftrms}) reduces to Eq.~(\ref{diffsion3}).

\section{$3\times 3$ matrices}
\label{app:3dimmatrix}

Let us consider the three coupled Langevin equations Eqs.~(\ref{eqr12}), (\ref{eqr23}), and (\ref{efflangevin}). 
By introducing the three-dimensional vectors ${\bm r} = (r_{12}, r_{23}, X)^{\mathsf T}$ and 
${\bm \xi} = (\xi_{12} , \xi_{23}, \xi_X)^{\mathsf T}$, 
the corresponding $3\times 3$ matrices ${\bf A}$, ${\bf F}$, and ${\bf D}$ are given by 
\begin{align}
{\mathbf A} = 
\begin{pmatrix} 
-K/\zeta_{12} &  K/\zeta_2 & 0
\\
K/\zeta_2 & -K/\zeta_{23} & 0 
\\
\dfrac{K}{3}\left( \dfrac{\zeta_{2} - \zeta_{1}}{\zeta_{1}\zeta_{2}} \right) & \dfrac{K}{3}\left( \dfrac{\zeta_{3} - \zeta_{2}}{\zeta_{2}\zeta_{3}} \right) & 0
\end{pmatrix},
\label{appAmatrixthree}
\end{align}
\begin{align}
{\mathbf F} = 
\begin{pmatrix} 
\left( \dfrac{2T_{12}}{\zeta_{12}} \right)^{1/2} &  0 & 0
\\
0 & \left( \dfrac{2T_{23}}{\zeta_{23}} \right)^{1/2} & 0
\\
0 & 0 & \dfrac{\sqrt{2}}{3}\left( \dfrac{T_{1}}{\zeta_{1}} + \dfrac{T_{2}}{\zeta_{2}} + \dfrac{T_{3}}{\zeta_{3}} \right)^{1/2} 
\end{pmatrix},
\label{appFmatrixthree}
\end{align}
and 
\begin{align}
{\bf D}  = \begin{pmatrix} 
T_{12}/\zeta_{12} &  -T_2/\zeta_2 & \dfrac{1}{3}\left( \dfrac{T_{2}}{\zeta_{2}} - \dfrac{T_{1}}{\zeta_{1}}\right)
\\
-T_2/\zeta_2 & T_{23}/\zeta_{23} & \dfrac{1}{3}\left( \dfrac{T_{3}}{\zeta_{3}} - \dfrac{T_{2}}{\zeta_{2}}\right) 
 \\
\dfrac{1}{3}\left( \dfrac{T_{2}}{\zeta_{2}} - \dfrac{T_{1}}{\zeta_{1}}\right) & \dfrac{1}{3}\left(  \dfrac{T_{3}}{\zeta_{3}} - \dfrac{T_{2}}{\zeta_{2}}\right)  & \dfrac{1}{9}\left(  \dfrac{T_{1}}{\zeta_{1}} + \dfrac{T_{2}}{\zeta_{2}} + \dfrac{T_{3}}{\zeta_{3}} \right)
\end{pmatrix},
\label{appDmatrixthree}
\end{align}
respectively.
The above matrices are the generalization of Eqs.~(\ref{Amatrixthree}), (\ref{Fmatrixthree}), 
and (\ref{Dmatrixthree}) to a higher dimension.

\end{widetext}


\end{document}